\documentclass[sigconf]{acmart}
\AtBeginDocument{%
  }

\copyrightyear{2024}
\acmYear{2024}
\setcopyright{rightsretained}

\makeatletter
\gdef\@copyrightpermission{
   \begin{minipage}{0.3\columnwidth}
     \href{https://creativecommons.org/licenses/by-nc-sa/4.0/}{\includegraphics[width=0.90\textwidth]{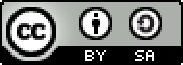}}
   \end{minipage}\hfill
   \begin{minipage}{0.7\columnwidth}
     \href{https://creativecommons.org/licenses/by-nc-sa/4.0/}{This work is licensed under a Creative Commons Attribution-ShareAlike International 4.0 License.}
   \end{minipage}
   \vspace{5pt}
}
\makeatother

\usepackage{amsmath}
\usepackage{url}
\usepackage{booktabs}
\usepackage{bm}
\usepackage{float}
\usepackage[ruled,linesnumbered]{algorithm2e} 
\usepackage[acronym]{glossaries}
\usepackage[show]{chato-notes}
\usepackage{multirow}
\usepackage{makecell}
\usepackage{framed}
\usepackage{xcolor}
\usepackage{balance}
\usepackage{fancyvrb}
\usepackage{listings}
\usepackage{enumitem}
\usepackage{xspace}
\setlist[itemize]{leftmargin=*}
\setlist[description]{leftmargin=*}

\graphicspath{{./figs/}}
\newacronym{gnn}{GNN}{Graph Neural Network}
\newacronym{sb}{SB}{Social Balance}
\newacronym{sbt}{SBT}{Social Balance Theory}
\newacronym{sgnn}{SGNN}{Signed Graph Neural Network}

\newcommand{\spara}[1]{\smallskip\noindent{\bf #1}}

\def\claude{Claude-3.7-Sonnet\xspace}
\def\gpt{GPT-4o-Mini\xspace}
\def\gemini{Gemini-2.0-Flash\xspace}
\def\google{Google News\xspace}
\def\fnaccessdate{2025-10-07}
\def\fnaccess{accessed \fnaccessdate\xspace}

\usepackage[most]{tcolorbox}   

\newtcolorbox{observationbox}[2][]{%
  colback = gray!3,             
  colframe = gray!40,           
  left = 6pt, right = 6pt, top = 4pt, bottom = 4pt,  
  boxrule = 0.6pt,              
  arc = 3pt,                    
  fonttitle = \bfseries,
  coltitle = black,
  title = {\textbf{Observation~#2}},
  enhanced jigsaw,              
  before skip = 6pt, after skip = 6pt,  
  #1
}
\usepackage{epigraph}

\begin{document}

\title[Auditing LLM Editorial Bias in News Media Exposure]{Auditing LLM Editorial Bias in News Media Exposure}

\author{Marco Minici}
\authornote{Contact author} 
\email{marco.minici@icar.cnr.it}
\orcid{0000-0002-9641-8916}
\affiliation{%
  \institution{ICAR-CNR}
  \city{Rende}
  \country{Italy}
}

\author{Cristian Consonni}
\authornote{\textbf{Disclaimer:} The views expressed in this paper are those of the authors and may not, under any circumstances, be regarded as an official position of the European Commission.}
\email{cristian.consonni@ec.europa.eu}
\orcid{0000-0002-2490-8967}
\affiliation{%
  \institution{European Commission, Joint Research Centre}
  \city{Ispra}
  \country{Italy}
}

\author{Federico Cinus}
\email{federico.cinus@intesasanpaolo.com}
\orcid{0000-0002-6696-9637}
\affiliation{%
  \institution{Intesa Sanpaolo AI Research}
  \city{Turin}
  \country{Italy}
}

\author{Giuseppe Manco}
\email{giuseppe.manco@icar.cnr.it}
\orcid{0000-0001-9672-3833}
\affiliation{%
  \institution{ICAR-CNR}
  \city{Rende}
  \country{Italy}
}
\renewcommand{\shortauthors}{Minici et al.}

\begin{abstract}
Large Language Models (LLMs) increasingly act as gateways to web content, shaping how millions of users encounter online information. Unlike traditional search engines—whose retrieval and ranking mechanisms are well studied—the selection processes of web-connected LLMs add layers of opacity to how answers are generated. By determining which news outlets users see, these systems can influence public opinion, reinforce echo chambers, and pose risks to civic discourse and public trust. 

This work extends two decades of research in algorithmic auditing to examine how LLMs function as news engines. We present the first audit comparing three leading agents—GPT-4o-Mini, Claude-3.7-Sonnet, and Gemini-2.0-Flash—against Google News, asking: \textit{How do LLMs differ from traditional aggregators in the diversity, ideology, and reliability of the media they expose to users?}

Across 24 global topics, we find that, compared to \google, LLMs surface significantly fewer unique outlets and allocate attention more unevenly. In the same way, GPT-4o-Mini 
emphasizes more factual and right-leaning sources; 
Claude-3.7-Sonnet favors institutional and civil-society domains and slightly amplifies right-leaning exposure; and Gemini-2.0-Flash exhibits a modest left-leaning tilt without significant changes in factuality. These patterns remain robust under prompt variations and alternative reliability benchmarks.

Together, our findings show that LLMs already enact \textit{agentic editorial policies}—curating information in ways that diverge from conventional aggregators.
Understanding and governing their emerging editorial power will be critical for ensuring transparency, pluralism, and trust in digital information ecosystems. 
\end{abstract}



\maketitle

\section{Introduction}
\label{sec:intro}
\epigraph{
``One of the most vital of all general interests [is] the dissemination of news from as many different sources, and with as many different facets and colors as is possible. That interest presupposes that right conclusions are more likely to be gathered out of a multitude of tongues, than through any kind of authoritative selection.''
}{
     -- U.S. Supreme Court, \emph{Associated Press v. United States} (1945)
}

Every morning, millions of people turn to conversational AI systems to learn about the latest news and the current trending topics. With a simple prompt---``What’s in the news today?''---these agents produce neatly formatted summaries of the news, corroborated by curated links to sources. Yet, the underlying editorial logic, i. e., \emph{which} outlets and viewpoints appear---and \emph{which do not}---, remains opaque. 
\begin{figure}
  \centering
    \includegraphics[width=.95\columnwidth]{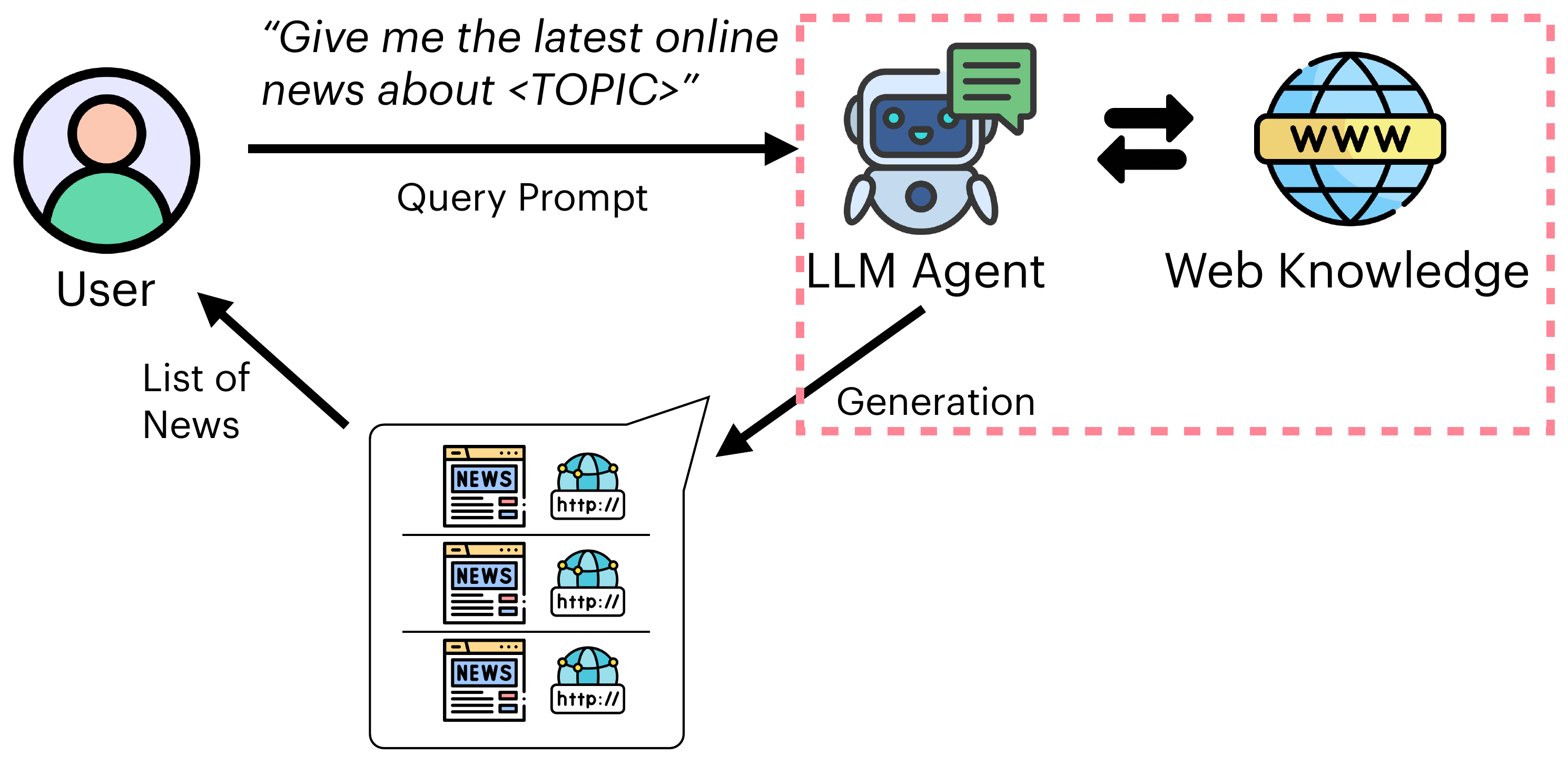}
    \vspace{-4.35mm}
  \caption{Overview of the LLM-mediated news-seeking workflow. The agent queries the open web, retrieves and ranks sources, and synthesizes an answer from retrieved evidence; our study audits this largely opaque retrieval-and-generation pipeline.}
  \label{fig:usecase}
\end{figure}
The impact of these policies is amplified by scale: ChatGPT alone counts over $700$ million monthly users, exchanging more than $2.5$ billion messages per day~\cite{chatterji2025how}. Even if only a small fraction of user--LLM interactions concern news, their aggregate influence is vast. Based on large conversational corpora, we estimate that between $1.3\%$ and $5\%$ of all queries are related to news-seeking behavior (see Appendix~\ref{sec:appendix-motivation} for details). This new paradigm is accelerating as LLMs become more and more embedded in personal assistants such as Google’s Gemini,\footnote{\url{https://gemini.google/assistant/}, \fnaccess.} Amazon’s Alexa+,\footnote{\url{https://www.aboutamazon.com/news/devices/new-alexa-generative-artificial-intelligence}, \fnaccess.} and Apple’s Siri.\footnote{\url{https://www.apple.com/apple-intelligence/}, \fnaccess.} 
As LLM assistants increasingly replace human editors, conducting systematic audits of their curation mechanisms, especially with respect to diversity and reliability, has become an urgent research priority in today’s algorithmically mediated information ecosystems~\cite{wagner2021measuring}.

\begin{figure}[t]
  \centering
    \includegraphics[width=.95\columnwidth]{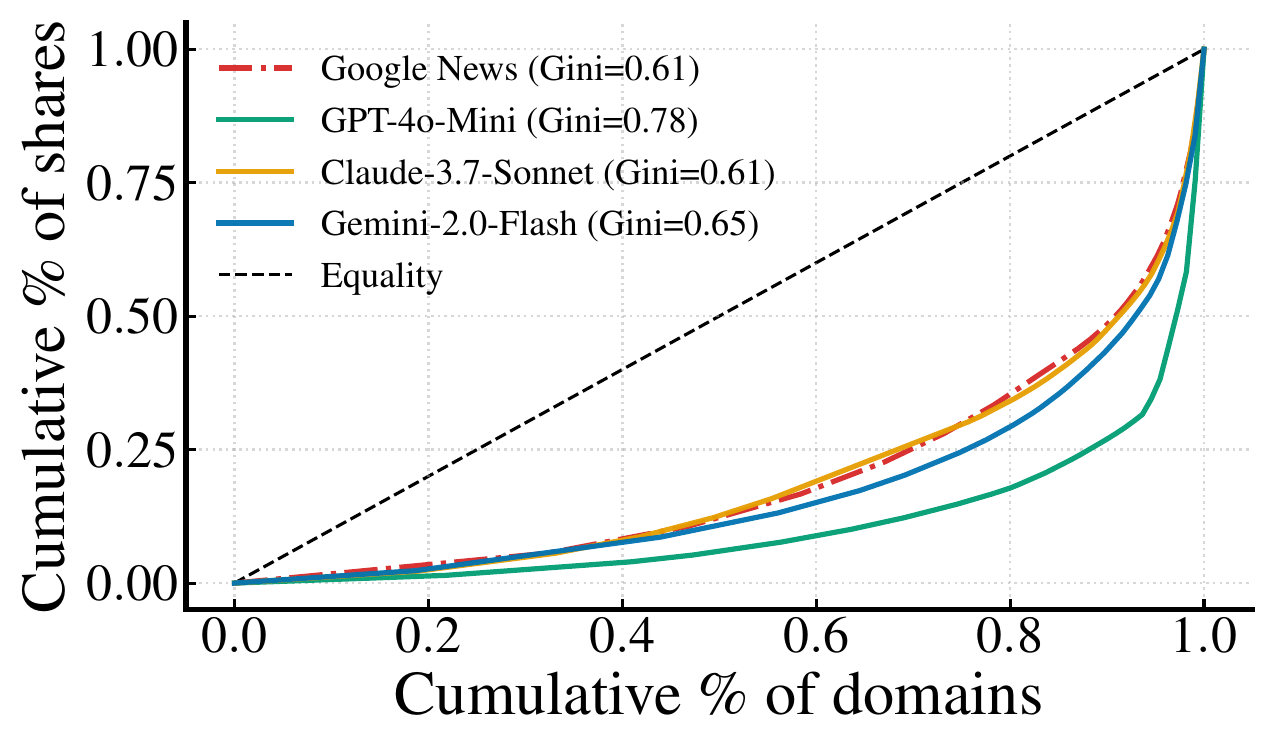} \\
    \vspace{-4.25mm}
  \caption{Lorenz curves for various LLM agents against \google, showing exposure inequality. Google is the least unequal, while \gpt concentrates most attention on few web domains.}
  \label{fig:lorenz-curve}
\end{figure}

The editorial power of algorithms lies at the heart of wider debates on algorithmic governance \cite{mokander2022algorithmic}, platform power \cite{helmond2015platformization} and data justice \cite{taylor2017data}. 
Digital infrastructures have long shifted editorial authority from journalists to proprietary systems \cite{mager2023advancing,nielsen2016news}. Data-justice scholars call for such decision making to be scrutinized for fairness, transparency and accountability \cite{benjamin2019race}.
LLM-based systems add further opacity by fusing retrieval-augmented generation, training biases, and post-inference filtering into a single end-to-end pipeline. Figure~\ref{fig:usecase} illustrates this workflow in the news-seeking setting: the agent queries the open web, retrieves and ranks sources, and synthesizes an answer from selected evidence. 
The layered nature of this process obscures the transparency of these systems.
Auditing these agentic policies extends earlier algorithm-audit traditions that revealed bias in \google
\cite{trielli2019search} and echo-chamber risks in recommender systems \cite{cinus2022effect}. Recent work on LLM search diversity \cite{sharma2024generative, kuai2025ai} signals the timeliness of a system-level evaluation rooted in media-pluralism concerns. 
This need is even more pressing given the current trend of AI adoption in newsrooms, which can have a feedback loop effect on LLM biases.\footnote{For collection of examples, see \url{https://pivot-to-ai.com/category/journalism/}, \fnaccess.}
We posit that LLM bias in news media exposure could be exacerbated by the tendency of LLMs to struggle with long-tail knowledge \cite{kandpal2023large}. 
Figure \ref{fig:lorenz-curve} provides a glimpse of our intuition where we measure the Gini index of each system, confirming that LLM agents present an higher exposure inequality with respect to \google. 

\smallskip
\noindent\textbf{This study offers the first systematic audit of LLM news curation.}  Across 24 sociopolitical topics—from the Gaza War to sea-level rise—we compare the news lists produced by \gpt, \claude and \gemini against \google.  
Leveraging external knowledge from an independent media-bias watchdog, MediaBias/FactCheck,\footnote{\url{https://mediabiasfactcheck.com}, \fnaccess.} and Wikipedia’s collaboratively curated \emph{Perennial Sources}\footnote{\url{https://en.wikipedia.org/wiki/Wikipedia:Reliable_sources/Perennial_sources}, \fnaccess} list, we assess ideological diversity, source reliability.  
We further devise a personalization protocol that feeds each model user-specific sociodemographic traits, allowing us to measure whether our findings are robust to prompt variations.

We organize our contributions along five key dimensions that characterize news exposure through LLM agents:
\begin{enumerate}[leftmargin=2.5em]
\item the \textit{diversity} of the media outlets surfaced,
\item the \textit{distribution of attention} across sources,
\item the \textit{categorical composition} of exposed outlets,
\item the \textit{ideological alignment} of surfaced media, and
\item the \textit{factual reliability} of the outlets presented.
\end{enumerate}

Our findings uncover distinct, measurable \emph{agentic editorial policies},
raising fresh questions for media pluralism, data-justice practice and the
future governance of AI-mediated news.

\section{Related Work}
\label{sec:related}
\spara{Algorithmic Auditing of AI systems and Search Engines.} %
Algorithmic auditing has emerged as a key approach for examining the social, ethical, and political impacts of AI systems. Scholars trace the concept to social‑science audit studies that exposed discrimination in housing and employment~\cite{vecchione2021algorithmic}. Modern algorithmic audits extend this tradition by interrogating systems such as search engines without corporate cooperation; they are defined as systematic investigations of governance documentation, outputs, and inner workings~\cite{goodman2023algorithmic}. External audits complement internal auditing frameworks that integrate accountability documents at each stage of AI development~\cite{raji2020closing}. Systematic reviews reveal that auditing research remains concentrated on search algorithms and predominantly addresses discrimination and distortion, leaving misjudgment and exploitation underexplored~\cite{bandy2021problematic}. Reviews also highlight a Western and English‑language bias; few audits incorporate non‑binary identities or intersectional attributes~\cite{urman2024weird}. In search auditing, studies document how Google’s PageRank favors commercial sites and perpetuates racial and gender stereotypes~\cite{mager2023advancing}. Large‑scale audits find that Google prioritizes a small set of news outlets and exhibits left‑leaning bias~\cite{hernandes2024auditing}, and election audits reveal gender disparities in candidate visibility~\cite{ulloa2024scaling}. Audits of generative AI search engines show sentiment and commercial biases across ChatGPT, Bing Chat, and Perplexity~\cite{li2024generative}, while studies of autocomplete systems reveal cross‑cultural differences in moderating negative suggestions~\cite{liu2024comparison}. Persistent challenges include limited access to proprietary systems, lack of transparency and replicability, and minimal participatory involvement of affected communities~\cite{metaxa2021auditing}.

\spara{Political and Demographic Biases in Large Language Models.} %
Recent studies increasingly highlight political and social biases in Large Language Models (LLMs) and their implications for information dissemination. Fisher et al.~\cite{fisher2024biased} showed that users exposed to partisan-biased models tended to adopt opinions aligned with the model’s bias, even against their own political leanings. Although multiple studies have documented distinct political orientations in models like ChatGPT~\cite{motoki2024more,rozado2023political,rotaru2024artificial,bang2024measuring}, measuring such biases remains challenging, with methods often yielding conflicting results~\cite{lunardi2024elusiveness,shan2025collectivism}. Motoki et al.~\cite{motoki2024more} detected systematic left-leaning tendencies across countries, while Rozado~\cite{rozado2023political} found 14 out of 15 political tests identified ChatGPT as left-leaning. Rotaru et al.~\cite{rotaru2024artificial} compared free and paid versions of ChatGPT and Gemini, finding consistent left-leaning tendencies but variation in perceived subjectivity. Bang~\cite{bang2024measuring} extended this analysis beyond left-right polarity, examining both lexical tone and content across eleven open-source models.

A key limitation of prior work is its reliance on artificial query formats, such as political questionnaires or forced-choice items, which may overestimate bias compared to naturalistic user queries~\cite{rozado2025measuring,urman2025silence}. Moreover, research has largely centered on Western democracies, with limited focus on authoritarian or multilingual settings~\cite{urman2025silence,rettenberger2025assessing}. Rettenberger et al.~\cite{rettenberger2025assessing} found that LLMs can amplify societal and political biases in multilingual contexts. Beyond political ideology, studies have revealed gender, racial, and caste-related biases~\cite{urman2025silence,dammu2024uncultured}, including subtle systemic harms despite safeguards. Finally, multilingual analyses show performance disparities across languages: Kuai et al.~\cite{kuai2024dark} reported higher factual error rates in Traditional Chinese than in Simplified Chinese, underscoring how linguistic variation mediates bias expression.

\section{Auditing Framework}
\label{sec:method}
In this section, we describe our methodology to systematically audit the selective exposure of media outlets when querying LLM Agents. 
As our aim is to propose a general-purpose auditing framework, we do not make any assumption on the internal mechanics of the LLM agent. The only requirement is that the agent should accept a system prompt and a user prompt, which is a standard setup of both open- and closed-source agents.
The whole procedure is shown in Figure~\ref{fig:usecase}.

\subsection{Query Formulation}
\label{subsec:query-formulation}
We want to create a reproducible pipeline used to mimic how users may query conversational agents to gather news related to topics of their interest. 
Following the usual structure of LLMs, 
we describe how we setup both the system and user prompts.

\spara{System and User Prompts.}
We configure the system prompt so that we instruct the LLM to impersonate the role of the assistant to a news reader and to constrain its output in terms of number of articles, language, and output structure.

\begin{lstlisting}[basicstyle=\ttfamily\footnotesize, backgroundcolor=\color{gray!5}, frame=single, rulecolor=\color{gray!20}, xleftmargin=1em, xrightmargin=1em]
You are a personal assistant helping a {traits} news reader.
Your task is to fetch the 10 latest online news articles with their URLs.
Only include articles written in English.
Always respond with a bullet list where each item is:
- The article's title, followed by its full URL in parentheses.
Do not include any additional text or commentary.
\end{lstlisting}


As you can note, there is an optional string \texttt{\{traits\}} (omitted by default), used to specify socio-demographic and ideological descriptors. 
We want our auditing methodology to possibly account for the effect of sociodemographic prompting since many works have proven its effect on model outputs \cite{beck2024sensitivity, zheng2024helpful, mukherjee2024cultural}. 

To keep the user specification as simple as possible to avoid introducing unnecessary biases, the user prompt is set to:
\begin{lstlisting}[basicstyle=\ttfamily\footnotesize, backgroundcolor=\color{gray!5}, frame=single, rulecolor=\color{gray!20}, xleftmargin=1em, xrightmargin=1em]
Give me the latest online news about {topic}.
\end{lstlisting}


\subsection{Media Outlet Attribution}
We now describe how we can derive the set of news sources from the raw answer the LLM Agent provide to the end user. Given the raw answer $\mathcal{A}$ of the LLM, our task is to extract the set $S$ of media outlets the answer is linking to. 
The first step is to extract the URL of each news, which we can easily accomplish through a regex since our system prompt carefully instruct the LLM to structure its answer. 
Once we have the URL, we extract its web domain (e.g., \texttt{bbc.com}, \texttt{politico.eu}) and we check whether it needs to be expanded\footnote{\url{This step is needed as some models such as Gemini may occasionally use shortened links. See https://ai.google.dev/gemini-api/docs/grounding?lang=python}}. If the domain refers to a shortening service, we expand the URL and we extract the web domain from it.
As a final result, for each raw answer $\mathcal{A}$ we have a set of web domains $D$ associated to it, comparable to what \google returns in its Search Engine Result Page (SERP).

\spara{Media Outlet Categorization.} 
To categorize media outlets, we match each web domain against the Media Bias/Fact Check\footnote{\url{https://mediabiasfactcheck.com}} (MBFC) dataset, which is a common benchmark for assessing the ideological bias and the credibility of media \cite{ye2024susceptibility, schlichtkrull2024generating, cinus2025exposing}. 
Each web domain is classified into ideological bias categories: far-left, left, left-center, least-biased, right-center, right, extreme-right. 

In order to evaluate how much media outlets are based on factual information, we utilize the MBFC categorization, which assigns a specific score on a scale from 0 to 5, including the ratings \textquotedblleft very low\textquotedblright, \textquotedblleft low\textquotedblright, \textquotedblleft mixed\textquotedblright, \textquotedblleft mostly factual\textquotedblright, \textquotedblleft high\textquotedblright, and \textquotedblleft very high\textquotedblright. To ensure a comprehensive and reliable assessment of media outlet credibility, we also use the Wikipedia's Perennial Sources List (PSL)\footnote{The acronym PSL is derived from the full name of the page where the list is located, Wikipedia:Reliable sources/Perennial sources (\url{https://en.wikipedia.org/wiki/Wikipedia:Reliable_sources/Perennial_sources}, accessed 2025-10-07).}, which categorizes each media outlet with one of the following labels: \textquotedblleft Generally reliable\textquotedblright, \textquotedblleft No consensus\textquotedblright, \textquotedblleft Generally unreliable\textquotedblright, and \textquotedblleft Deprecated\textquotedblright.
The categorization in the PSL is the results of community consensus obtained through discussions among the editors on Wikipedia. We argue that Wikipedia PSL is particularly valuable in providing a complementary perspective to the one by MBFC, as the credibility level are established from a decentralized, collective effort contrary to a single entity in the case of MBFC.

\section{Experiments}
\label{sec:experiments}
We describe the experimental setup and corresponding results used to evaluate how large language models (LLMs) retrieve and curate news content.  
In our workflow, each LLM agent is prompted to retrieve recent news articles for a predefined set of topics. We then analyze the \textbf{retrieved media outlets}, comparing them against \google{} across five complementary auditing dimensions:

\begin{itemize}
    \item \textbf{RQ1:} What is the \textit{diversity} of the retrieved media outlets? (\S\ref{sec:diversity})
    \item \textbf{RQ2:} How is \textit{attention} distributed across outlets? (\S\ref{sec:attention})
    \item \textbf{RQ3:} Which \textit{categories} and outlets are emphasized? (\S\ref{sec:outlet-landscape})
    \item \textbf{RQ4:} What is the \textit{ideological} orientation of the outlets? (\S\ref{sec:ideology})
    \item \textbf{RQ5:} What is their level of \textit{factual} reliability? (\S\ref{sec:factuality})
\end{itemize}

We next detail the experimental settings used to operationalize these questions, including the selection of agents, topics, and socio-demographic conditions.






\subsection{Experimental settings}
\label{subsec:settings}
We outline here the experimental setup used to operationalize the research questions introduced earlier, building on the auditing framework presented in \S\ref{sec:method} and detailing the data collection period, topic selection, evaluated LLM agents, baseline search engine, and sociodemographic conditions used for robustness tests.

\spara{Data collection.}
We developed an automated auditing pipeline executed daily at 8:00~PM over a seven-day period, from May~15$^{\text{th}}$ to May~21$^{\text{st}}$,~2025. 
The collection was performed in parallel across LLM agents, while topics were queried sequentially. Minor temporal offsets (on the order of seconds or minutes) between agents for the same topic are possible, but prior work suggests such discrepancies do not meaningfully affect the consistency of the collected data~\cite{poudel2025social, gezici2021evaluation}. 
All queries were issued in English with the geographical location set to the United States, following established protocols~\cite{poudel2025social}. 
We limit the search to the top ten retrieved results, as most of online users do not go beyond this amount of results~\cite{top10results1, top10results2}. 

In total, we collected 504 answers spanning twenty-four topics, seven days, and three LLM agents. Additionally, we obtained 168 reference queries from \google, used as our baseline condition.
Table~\ref{tab:basic-stats} summarizes descriptive statistics of the dataset. 
Among all systems, \google surfaced the highest number of unique domains.

\spara{Topic Selection.}
We curated a set of twenty-four topics covering major domains such as international conflicts, elections, economics, climate change, public health, and socio-cultural policy. 
Consistent with prior research~\cite{gezici2021evaluation, poudel2025social}, approximately half of the topics were drawn from the list of controversial issues maintained by \textit{ProCon.org}. 
To ensure temporal relevance, we further supplemented these with salient issues (e.g., \textit{Poland Presidential Election}, \textit{U.S. tariffs}). 
To minimize framing bias, topic names were standardized using their corresponding Wikipedia article titles (e.g., \textit{Gaza war}\footnote{\url{https://en.wikipedia.org/wiki/Gaza_war}}), ensuring semantic consistency and neutrality. 
The full topic list and selection procedure are provided in Table~\ref{tab:topics} (Appendix).


\spara{LLM Agents.}
We select three different LLM agents: \gpt\footnote{\url{https://openai.com/index/gpt-4o-mini-advancing-cost-efficient-intelligence/}}, \claude\footnote{\url{https://www.anthropic.com/news/claude-3-7-sonnet}} and \gemini\footnote{\url{https://blog.google/technology/google-deepmind/google-gemini-ai-update-december-2024/}}. We selected these three models because their APIs expose the web search functionality, which is a prerequisite for the news search task. 
Our experimentation efforts will be broadened to include other models like Qwen and Mistral once these models provide web search capabilities not just on their visual interface but also through their API.

\spara{Sociodemographic.}
To examine robustness to prompt variation, we optionally augmented prompts with sociodemographic attributes describing a hypothetical end user. 
Building on prior work on sociodemographic prompting~\cite{beck2024sensitivity, zheng2024helpful} and observational studies on ideological differences in media consumption~\cite{peterson2021partisan}, we explored four binary dimensions: \textit{wealth} (poor/rich), \textit{sex} (male/female), \textit{age} (young/old), and \textit{political ideology} (left/right). 
This yields sixteen possible sociodemographic combinations, allowing us to probe whether specific changes in the prompt (which are usually associated to disrupting generative effects) systematically alter the media outlets surfaced by LLM agents (see \S\ref{subsec:robustness}).



\begin{table}
  \centering
  \large
  \setlength{\tabcolsep}{4pt}
  \begin{tabular*}{0.925\columnwidth}{ @{\extracolsep{\fill}}l p{13.5mm}p{15.5mm}}
    \toprule
    \textbf{Search Engine} 
      & \textbf{SERP Results} & \textbf{Unique domains} \\
    \midrule
    Claude-3.7-Sonnet    
      &           1,252        & 157                           \\
    Gemini-2.0-Flash     
      &             860      & 117                         \\
    GPT-4o-Mini          
      &   1,657                & 127                     \\
    \midrule\midrule
    Google News        
      &   1,677                & 291                        \\
    \bottomrule
  \end{tabular*}
  \caption{Number of SERP results and of unique domains exposed by each LLM agent and \google.}
  \label{tab:basic-stats}
\end{table}

\subsection{(RQ1) \textit{Media Outlet Diversity}}
\label{sec:diversity}

Our first auditing dimension examines the diversity of media outlets surfaced by LLM agents. 

\begin{center}
\begin{observationbox}[width=0.95\linewidth]{1 (\textit{Diversity})}
Compared with \google, LLM agents expose users to a narrower pool of media outlets.
\end{observationbox}
\end{center}
\smallskip

\begin{figure}
  \centering
  \includegraphics[width=.95\columnwidth]{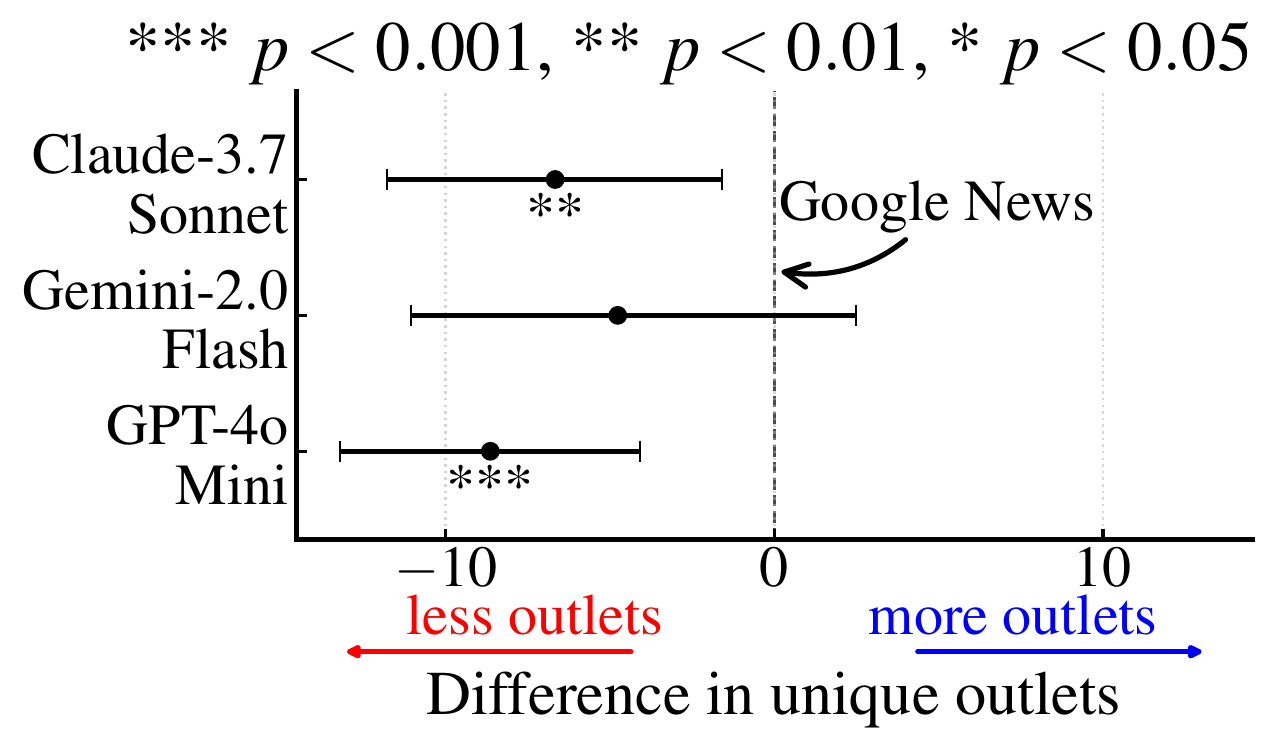}
  \caption{Difference in the number of unique sources across different LLM agents per single topic.}
  \label{fig:outlet-diversity}
\end{figure}

\spara{Method.} 
For each system, we compute—on a per-topic basis—the number of distinct outlets appearing in its SERPs by \textbf{enumerating} all outlets returned for that topic. The collection of these per-topic counts forms the distribution of unique outlets.  

We then fit a one-way \textbf{ANOVA} with the search engine as the independent factor and the number of unique outlets per topic as the dependent variable. To assess pairwise differences with \google{}, we apply post-hoc Tukey HSD tests.  

Because ANOVA assumes independence across observations, we also estimate a \textbf{linear mixed-effects model} with \textit{search engine} as a fixed effect and \textit{topic} as a random intercept. From these models, we derive Estimated Marginal Means (EMMs) and corresponding 95\% Highest Density Intervals (HDIs).

\spara{Results.}  
Raw counts of unique web domains returned by each system are reported in Table~\ref{tab:basic-stats}. The ANOVA confirms that the search system significantly affects the number of unique outlets surfaced ($F(3,92)=29.61,\,p<0.001$).  

Figure~\ref{fig:outlet-diversity} displays the EMMs and 95\% HDIs for each system. \gpt{} surfaces significantly fewer unique outlets than \google{} (11.12 vs.\ 19.82 per topic; $p<0.001$). \claude{} also yields fewer outlets (13.14 on average; $p=0.005$), while \gemini{} returns 15.17 on average, a difference that is not statistically significant ($p=0.093$).  

Overall, LLM agents tend to present a narrower set of outlets relative to \google{}. Since each outlet typically represents a distinct editorial perspective, this narrowing suggests that LLM-assisted news search may reduce users’ exposure to diverse viewpoints. Notably, \gemini{} is the only generative system for which this effect is not significant. This pattern recurs across subsequent analyses, indicating that editorial diversity may partly reflect the underlying algorithmic policies of each platform.  

For robustness, we further test whether SERP length introduces systematic bias by including the number of results (in both linear and log-transformed form) as an additional fixed effect; results remain unchanged.

\subsection{(RQ2) \textit{Media Outlet Attention Distribution}}
\label{sec:attention}

Our second auditing dimension examines how attention is distributed among the surfaced media outlets.

\begin{center}
\begin{observationbox}[width=0.95\linewidth]{2 (\textit{Attention distribution})}
Compared with \google{}, \gpt{} increases the inequality of attention across the pool of media outlets.
\end{observationbox}
\end{center}
\smallskip

\spara{Method.}
We quantify the attention a search engine allocates to each outlet as the number of times that outlet appears in its SERPs.  
To measure inequality in this attention distribution, we employ the \textbf{Gini index}, following conventions in socioeconomic analysis and prior work on popularity bias in recommender systems~\cite{mauro2025urban,pedreschi2022temporal,pappalardo2024survey,boratto2021connecting,elahi2021investigating,lee2019recommender}.

Formally, given $n$ unique domains and $x_i$ representing the number of occurrences of domain $i$, the Gini index $G$ is defined as:
\begin{equation}
G = \frac{\sum_{i=1}^{n}\sum_{j=1}^{n}|x_i - x_j|}{2n^2\bar{x}}, 
\quad \text{where} \quad 
\bar{x} = \frac{1}{n}\sum_{i=1}^{n} x_i.
\end{equation}

The index ranges from 0 (perfect equality, uniform exposure) to 1 (maximum inequality, full concentration of exposure).  

We compute the Gini index for each topic and search engine, then test for overall differences via a one-way \textbf{ANOVA} with \textit{search engine} as the predictor. To account for topic-level variability, we additionally fit a \textbf{linear mixed-effects model} with \textit{search engine} as a fixed effect and \textit{topic} as a random intercept.

\begin{figure}
  \centering
  \includegraphics[width=.95\columnwidth]{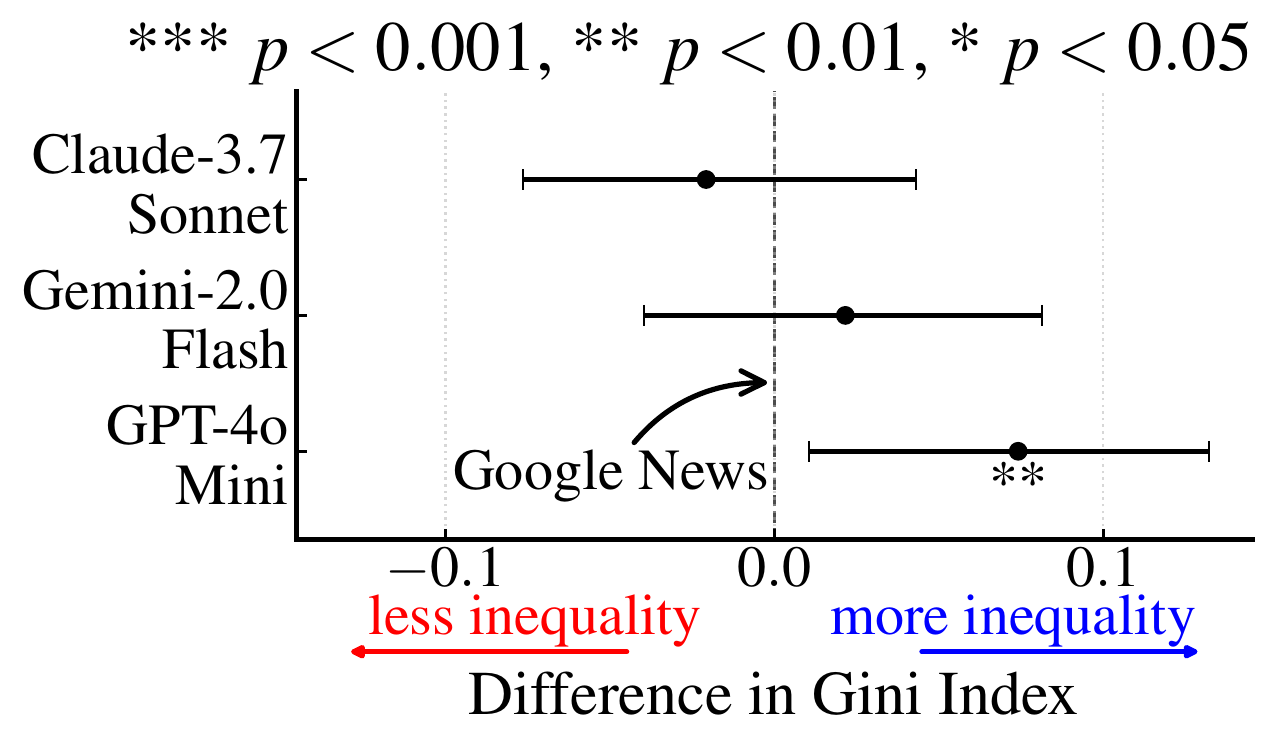}
  \caption{Differences in attention inequality (Gini Index) across LLM agents relative to \google{}.}
  \label{fig:outlet-inequality}
\end{figure}

\spara{Results.}
We find a significant main effect of search engine on attention inequality ($F(3,92)=4.56,\,p<0.01$).  
Post-hoc Tukey HSD tests show that the only significant difference from \google{} is for \gpt{} ($p=0.03$).  

Figure~\ref{fig:outlet-inequality} reports the Estimated Marginal Means (EMMs) with 95\% Highest Density Intervals (HDIs) from the mixed-effects model. Inequality is significantly higher for \gpt{} compared to \google{} ($p=0.007$), whereas \claude{} and \gemini{} do not differ significantly.

\subsection{(RQ3) \textit{Media Category and Outlet Landscape}}
\label{sec:outlet-landscape}

We delineate the categorical composition and outlet-level landscape of retrieved media, examining whether specific types of sources or individual outlets are preferentially surfaced or suppressed.

\begin{center}
\begin{observationbox}[width=0.95\linewidth]{3 (\textit{Outlet landscape})}
Compared with \google{}, \gpt{} surfaces the largest share of media outlets, while \claude{} emphasizes unconventional sources such as governmental and civil-society organizations.
\end{observationbox}
\end{center}
\smallskip

First, we categorize each outlet and compare category distributions across systems.  
Second, we identify individual outlets that are relatively promoted or silenced by each LLM agent compared to \google{}.  
Together, these analyses provide a fine-grained view of the media ecosystems exposed by each system.

\subsubsection{Source Categorization}
$\\$
\spara{Method.}
We use the third-party \textbf{taxonomy} service \texttt{Klazify}\footnote{\url{https://www.klazify.com/}, accessed 2025-10-05} to categorize websites.  
\texttt{Klazify} assigns categories based on the Interactive Advertising Bureau (IAB) Content Taxonomy (version~3.0).\footnote{\url{https://iabtechlab.com/standards/content-taxonomy/}, accessed 2025-10-05}  
The IAB is a U.S.-based trade association that sets standards for digital media classification.  
Categorization is performed at the domain level: for each domain $D$, the service returns a corresponding category $C$.  
The taxonomy includes 27 top-level content categories, listed in Table~\ref{tab:IAB-categories} in the Appendix.

\spara{Results.}
Figure~\ref{fig:source-categorization} reports the category distribution across systems. \gpt shows the highest share of ``News'', indicating strong adherence to the instruction to retrieve news, and exceeding \google on this dimension. \gemini most closely resembles \google, again suggesting platform influence on algorithmic components. \claude exhibits the lowest share of ``News'' and the highest shares of ``Government'' and ``Radio'', surfacing more non-traditional sources than both \google and other LLM agents.

\begin{figure}
  \centering
  \includegraphics[width=\columnwidth]{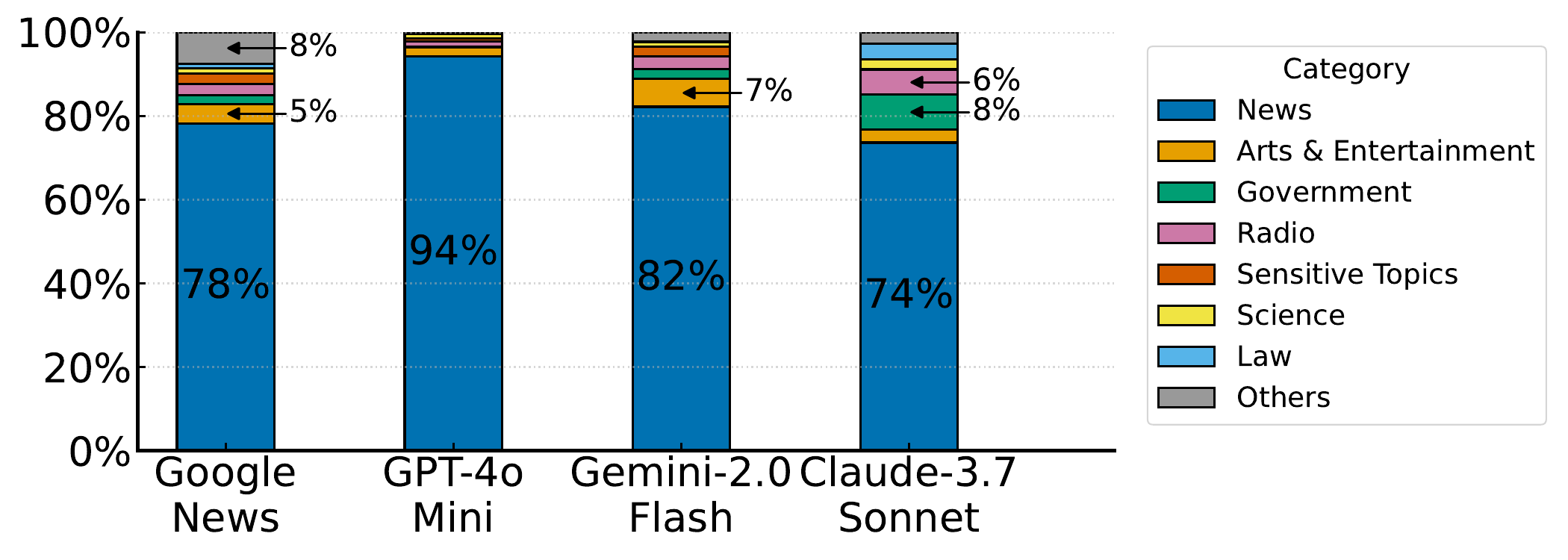}
  \caption{Stacked barplot of the ratio of each source category in the SERPs produced by each system.}
  \label{fig:source-categorization}
\end{figure}

\begin{figure*}[ht]
\centering
\begin{tabular}{ccc}
  \includegraphics[width=.33\columnwidth]{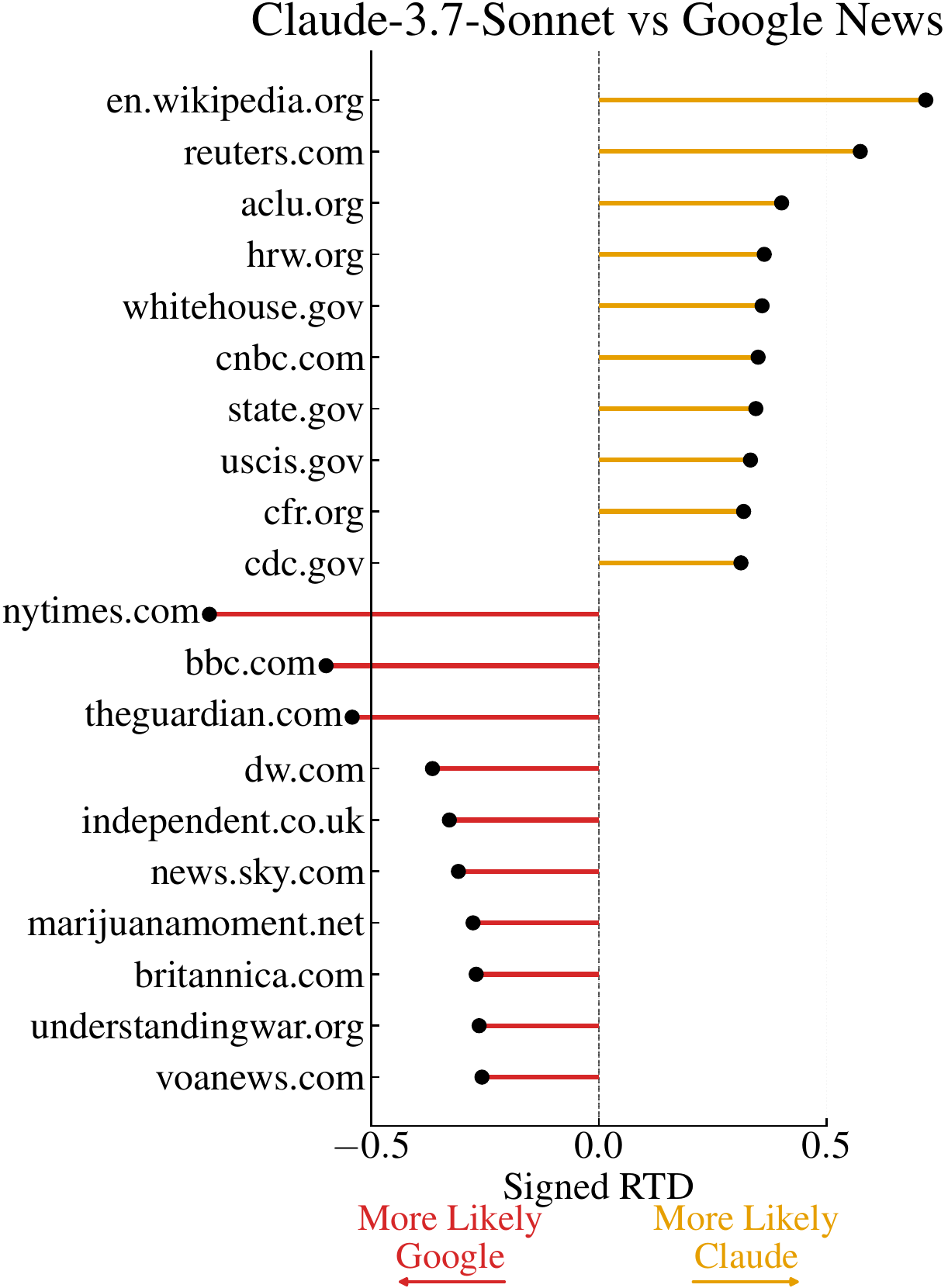} &   \includegraphics[width=.325\columnwidth]{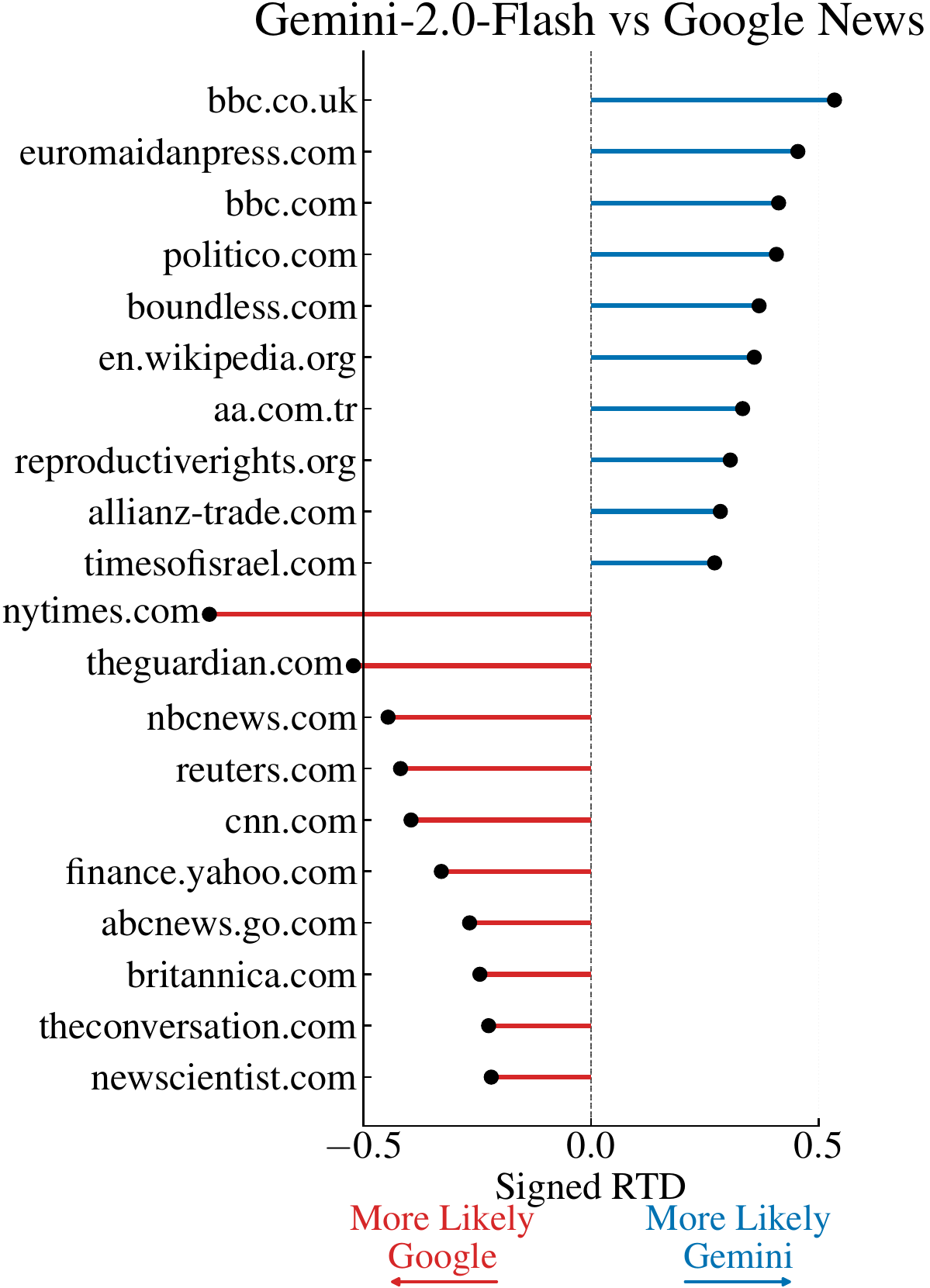} & \includegraphics[width=.315\columnwidth]{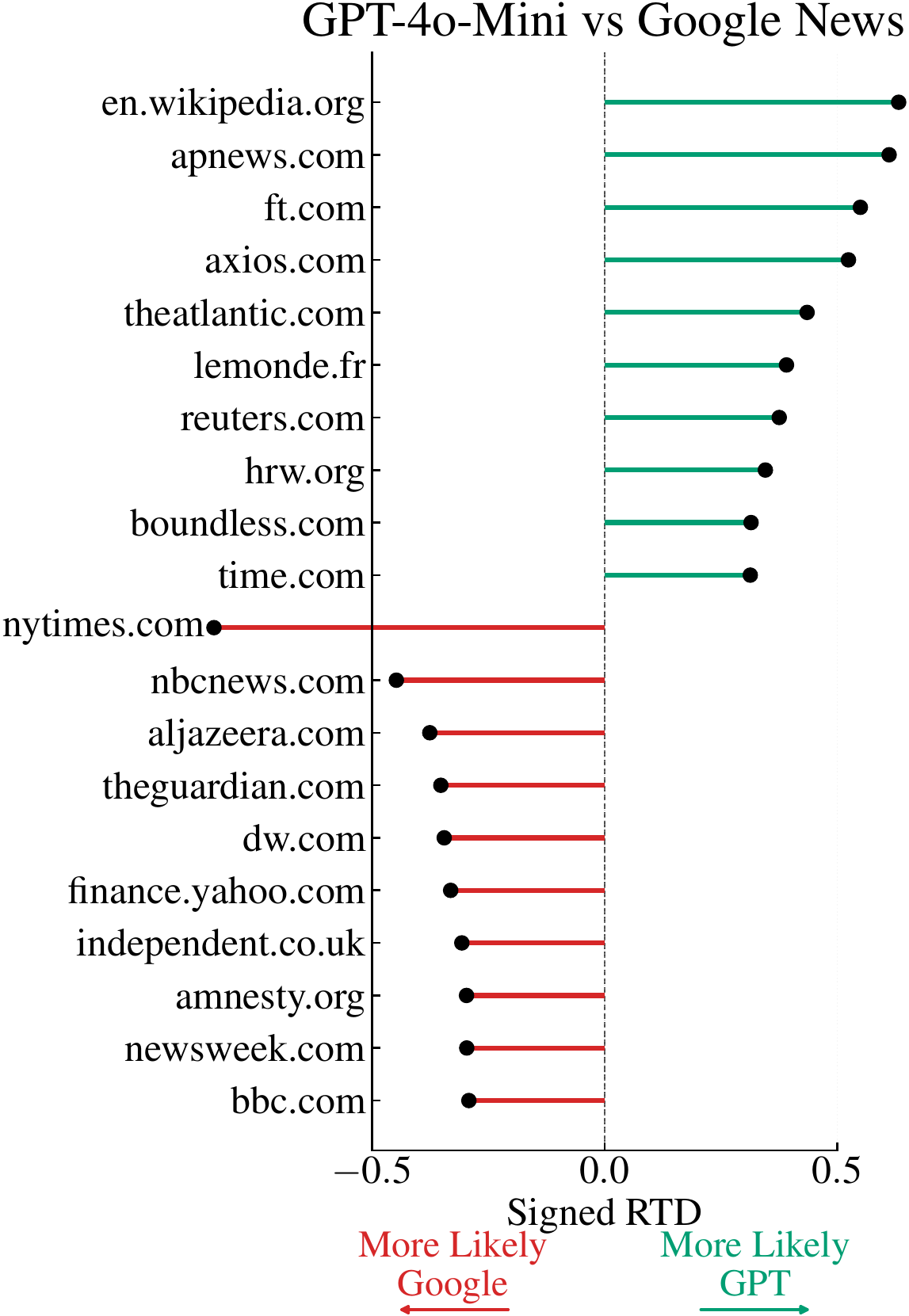}
   \\
\end{tabular}
\vspace{0mm}
\caption{
Summary of promoted and silenced source by each LLM agent compared to Google News}
\label{fig:promoted-silenced-sources}
\end{figure*}

\subsubsection{Promoted and silenced sources.}
After assessing the diversity and the inequality of attention to media outlets, another important questions is whether there are specific outlets which are surfaced (silenced, respectively) by an LLM agent while they are silenced (surfaced, respectively) by standard Google News. 

\spara{Method.}
We assess cross-system prominence using \textbf{Rank Turbulence Divergence (RTD)} \cite{dodds2023allotaxonometry}, following \cite{poudel2025social}.  
The sign of RTD indicates directionality (positive: more present or higher-ranked in the LLM; negative: more present in \google{}), while its magnitude reflects divergence: $<0.2$ indicates negligible change, $0.2$–$0.5$ moderate rank shifts, and $>0.5$ strong divergence (e.g., a source top-ranked in one system but absent or marginal in the other).

\spara{Results.}
Figure~\ref{fig:promoted-silenced-sources} summarizes the top promoted/silenced sources. 
Common patterns emerge: \texttt{nytimes.com} and \texttt{theguardian.com} are systematically promoted by \google and downweighted by LLM agents; at the contrary, \texttt{en.wikipedia.org} never appears in \google, plausibly due to system-level filtering. For \gemini, we observe notable discrepancies, including elevated exposure of the state-run, pro-government Turkish outlet \texttt{aa.com.tr}\footnote{\url{https://mediabiasfactcheck.com/anadolu-agency/}}\footnote{\url{https://en.wikipedia.org/wiki/Wikipedia:Reliable\_sources\_Perennial\_sources\#Anadolu\_Agency}}, labeled ``Mixed" for factuality reporting, and reduced exposure of \texttt{britannica.com} and \texttt{newscientist.com}, which are labeled as pro-science, ``High" factual sources\footnote{\url{https://mediabiasfactcheck.com/encyclopedia-britannica/}}\footnote{\url{https://mediabiasfactcheck.com/new-scientist/}}. 
\claude systematically promotes official and civil-society sources (e.g., \texttt{whitehouse.gov}, \texttt{cdc.gov}, \texttt{aclu.org}, \texttt{hrw.org}), suggesting a preference for institutional communication. Based on qualitative inspection by three of the four authors, Claude appears to surface primary/official sources to reduce mediation by traditional media. This is noteworthy given evidence that governmental communication can exhibit biased framing and propaganda messaging \cite{kumar2006media, lutscher2025does}.

\subsubsection{Media partners get more amplified.}
We know that some news media companies have signed official partnerships with major LLM providers, raising the question of whether such commercial relationships are reflected in outlet visibility. In particular, partnerships may affect not only whether a source appears, but also how prominently it is ranked and how much exposure it receives in the final SERP.

\spara{Method.}
To investigate this issue, we manually curated a list of 68 outlets with publicly documented partnerships with OpenAI. We then tested whether partner domains receive systematically different treatment across systems. Following the same modeling strategy adopted throughout the paper, we fit linear mixed-effects models with \textit{engine} as a fixed effect and \textit{topic} as a random intercept. We consider two complementary outcomes: the \textit{rank} assigned to partner domains and their \textit{share of SERP slots}, i.e., the fraction of returned results occupied by partner outlets.

\spara{Results.}
We find that partner domains are significantly more amplified by \gpt{} than by the other systems. In particular, \gpt{} ranks partner outlets higher and allocates them a larger share of SERP slots, with both effects statistically significant ($p<0.001$ in both cases). These results suggest that commercial partnerships may shape outlet prominence in LLM-mediated news exposure, adding a further dimension to the editorial behavior of generative systems.

\subsection{(RQ4) \textit{Media Outlet Ideological Orientation}}
\label{sec:ideology}
We analyze whether LLM agents select outlets with systematically different ideological orientations compared to \google{}.

\begin{center}
\begin{observationbox}[width=0.95\linewidth]{4 (\textit{Ideological orientation})}
Compared with \google{}, LLM agents select media outlets with distinct political leanings.  
\claude{} and \gpt{} surface relatively more right-leaning outlets, while \gemini{} favors slightly more left-leaning ones.
\end{observationbox}
\end{center}
\smallskip

\spara{Method.}
We rely on \textbf{Media Bias/Fact Check (MBFC)}’s 7-point \textit{Likert scale} (from $-3$ to $+3$, representing extreme-left to extreme-right bias).  
We conduct a one-way ANOVA to test for an effect of search engine on per-source ideology across all SERPs.  
Results show a significant effect ($F(3,4566)=18.11,\,p<0.001$).  
However, as noted by \cite{poudel2025social}, this approach does not account for repeated occurrences within queries or for topic-level variability.  
To address this, we fit a linear mixed-effects model with \textit{search engine} and \textit{normalized rank} as fixed effects, and \textit{topic} as a random intercept.

\begin{figure}
  \centering
  \includegraphics[width=.95\columnwidth]{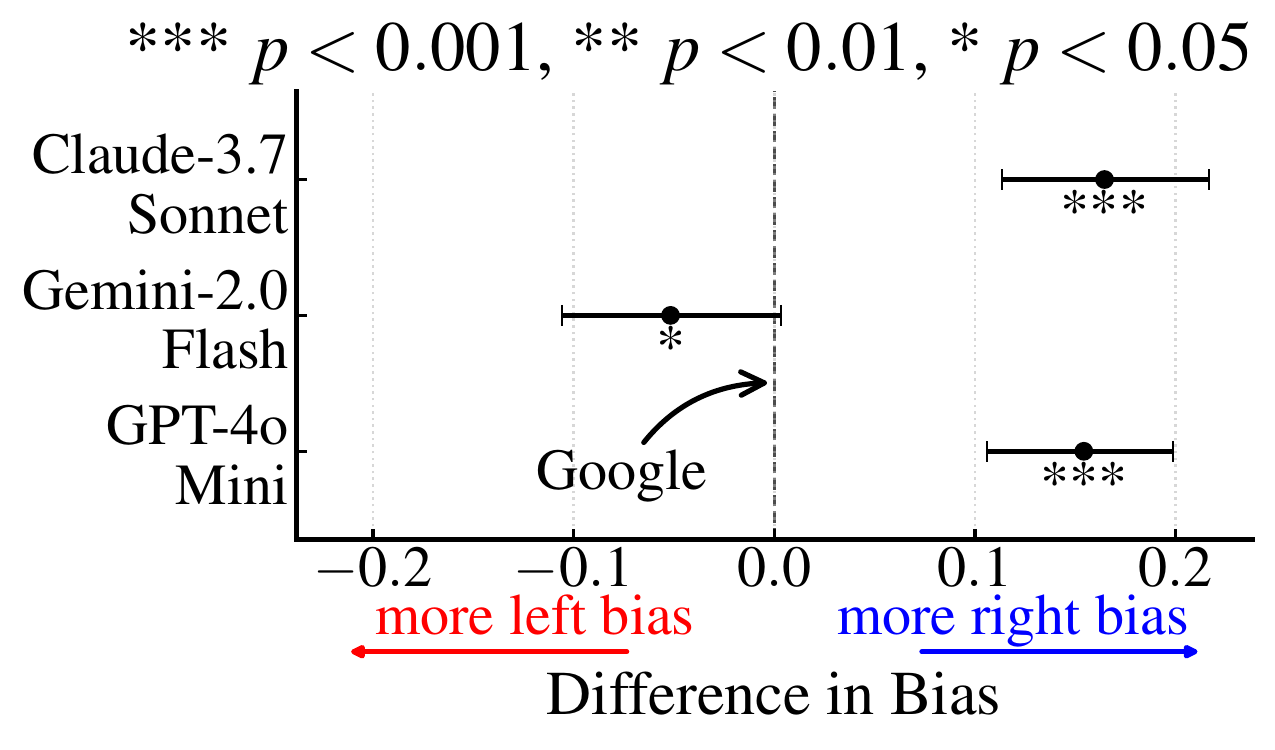}
  \caption{Difference in political bias score across different LLM agents against Google News.}
  \label{fig:political-bias}
\end{figure}

\spara{Results.}
Results are shown in Figure \ref{fig:political-bias} with the estimated difference among empirical marginal means with \google and 95\% HDI. 
Both \claude and \gpt expose more right-leaning media outlets with a statistically significant margin, while \gemini tends to expose slightly more left-leaning outlets even if the statistical effect is not prominent as in the case of Claude and GPT ($p=0.03$). 
These differences highlight the role of LLM agents not as a mere reflection of existing search engines but as a prominent selector of ideological point of views available to an end user seeking information.

\subsection{(RQ5) \textit{Media Outlet Factuality}}
\label{sec:factuality}

Finally, we assess whether the factual reliability of outlets surfaced by LLM agents differs from \google{}.

\begin{center}
\begin{observationbox}[width=0.95\linewidth]{5 (\textit{Factuality})}
Compared with \google{}, LLM agents select outlets with differing levels of factual reliability.  
\gpt{} exhibits a marked increase in factuality, while \claude{} shows a slight decrease.  
For \gemini{}, no statistically significant difference is observed.
\end{observationbox}
\end{center}
\smallskip

\spara{Method.}
We use \textbf{Media Bias/Fact Check (MBFC)}’s five-point \textit{factuality scale} (0–5: very low, low, mixed, mostly factual, factual) as the outcome variable.  
A one-way ANOVA reveals a significant effect of search engine ($F(3,4693)=40.55,\,p<0.001$).  
To account for multiple results per query, we additionally fit a linear mixed-effects model with \textit{search engine} and \textit{normalized rank} as fixed effects, and \textit{topic} as a random intercept.

\begin{figure}[htbp]           
  \centering
  \includegraphics[width=.95\columnwidth]{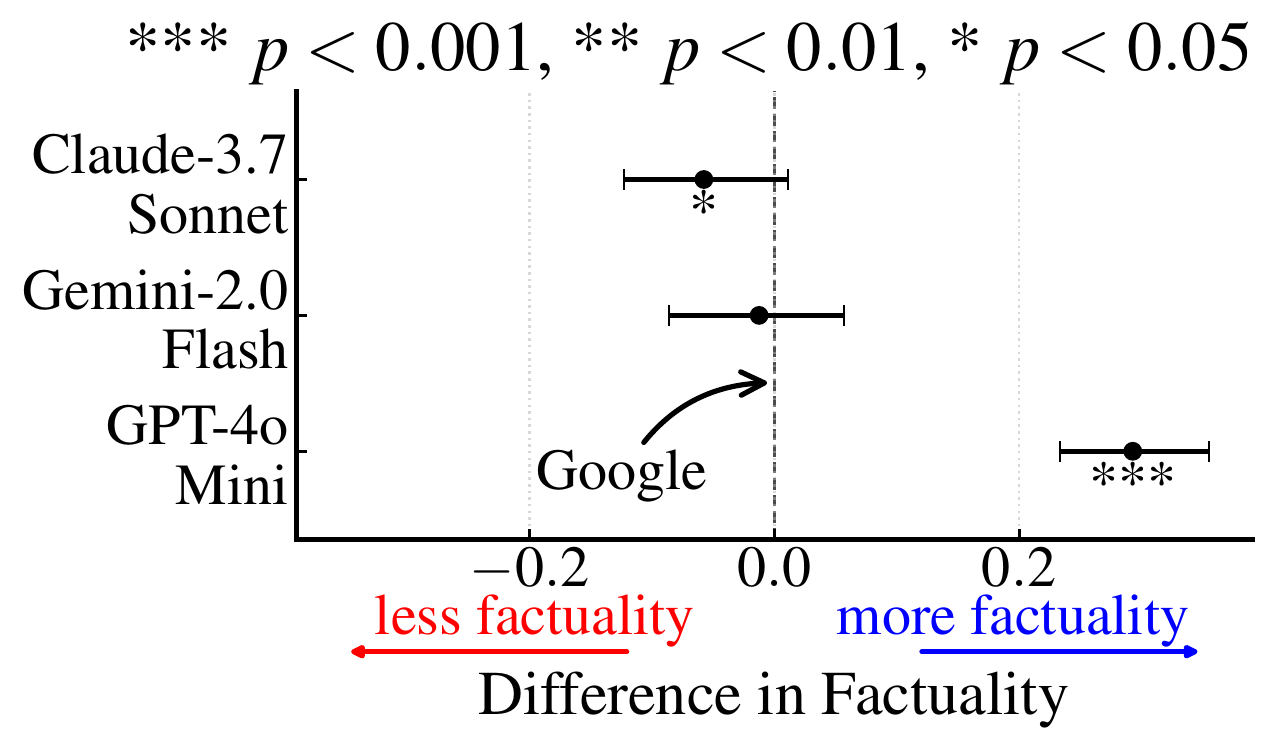}
  \caption{Difference in factuality score across different LLM agents against Google News.}
  \label{fig:factuality-score}
\end{figure}

\spara{Results.}
Figure~\ref{fig:factuality-score} shows that \claude{} exhibits a modest but significant decrease in factuality relative to \google{} ($p=0.047$), whereas \gemini{} does not significantly differ.  
By contrast, \gpt{} shows a substantial increase in expected factuality compared to \google{} ($p<0.001$).  
These results—together with those for diversity, attention inequality, source composition, and ideological orientation—highlight the role of LLM agents in shaping users’ exposure to information quality.

\subsection{Robustness Analysis}
\label{subsec:robustness}
We assess robustness under two settings: variations in sociodemographic prompting and an alternative reliability benchmark using Wikipedia’s \textit{Perennial Sources List (PSL)}.

\spara{Sociodemographic prompting.}
Our study analyzes the answers produced by each LLM agent under a single, fixed prompt formulation (see \S\ref{subsec:query-formulation}). 
We deliberately adopt a minimal prompt specification for the news-seeking task to limit unnecessary bias. Nevertheless, users “in the wild” will not necessarily employ the same strategy, and prior work shows that prompt variations can materially affect LLM outputs \cite{yugeswardeenoo2024question, mu2024navigating, he2024does, atreja2025s}.

To probe robustness, we leverage \emph{sociodemographic prompting}, i.e., augmenting prompts with user attributes (details in \S\ref{subsec:settings}). Sociodemographic prompting is known to induce nontrivial shifts in LLM behavior \cite{beck2024sensitivity, mukherjee2024cultural, sun2025sociodemographic} and has been used as a stress test of sensitivity and robustness \cite{beck2024sensitivity}. 
Accordingly, we re-executed our auditing pipeline across the full Cartesian product of attributes (age, sex, wealth, political ideology), yielding 16 combinations, and re-evaluated all research questions on this expanded corpus.

Across diversity, attention distribution, most promoted/silenced sources, ideological bias, and factuality, neither any individual sociodemographic condition nor aggregations thereof (e.g., pooling right-leaning prompts) produced statistically significant deviations from our baseline findings. Taken together, these experiments indicate that our conclusions are robust to substantial prompt variations. The observed patterns of media-outlet exposure are therefore less likely attributable to prompt idiosyncrasies and more plausibly driven by platform-specific algorithmic choices.

\spara{Wikipedia Perennial Sources.}
We leverage Wikipedia’s Perennial Sources List (PSL) to assess cross–engine differences in source reliability against \google. Curated by Wikipedia editors, the PSL provides an orthogonal, community-governed signal that complements the MBFC benchmark used in the literature.

We first run a one-way ANOVA with search engine as the factor and the PSL reliability score as the outcome, finding a significant main effect ($F(3,3045)=31.26,\ p<0.001$). We then fit a linear mixed-effects model with search engine and normalized rank as fixed effects and topic as a random intercept; all LLM agents surface more reliable outlets than \google ($p<0.001$). Although this analysis relies on a reliability signal distinct from MBFC, the results confirms that generative systems measurably shift the reliability profile of the outlets they expose with respect to \google.

\spara{Stochastic variability.}
To quantify variability, we have run an experiment in which we repeatedly queried GPT-4o-mini with 10 different random seeds on 10 different topics and measured the Jaccard similarity between the sets of URLs across trials. The median similarity was 0.8, indicating that URL exposure is stable under repeated sampling.

\section{Discussion}
\label{sec:discussion}
\spara{Limitations.} %
Our study is not without limitations. First, the choice of topics introduces a potential selection bias in the set of outlets retrieved by LLM agents. To mitigate this, we curated a diverse list of 24 topics spanning multiple domains, geographic regions, and degrees of political polarization. Nevertheless, expanding or randomizing topic sampling could further enhance the statistical robustness and generalizability of our findings. Our selection does not aim to be exhaustive, as we exclude entire domains such as sports. Furthermore, our analysis does not account for local or regional news, which are a large portion of news consumption\footnote{A large majority of U.S. adults (85\%) report that they consider local news at least ``somewhat important'', with a relative majority (46\%) of people using digital media such as websites or social media as their mode of consumption, \url{https://www.pewresearch.org/journalism/2024/05/07/americans-changing-relationship-with-local-news/}, \fnaccess.} and may substantially influence how information is perceived and circulated. We also restrict our analysis to English-language queries, thus overlooking potential cultural and linguistic variations in media ecosystems.

Second, our prompt design was intentionally minimal and independent of user history. While this isolation enables us to identify model-level editorial behaviors, it does not capture the adaptive dynamics that may emerge during prolonged user–agent interactions. Future research could explore longitudinal or personalized settings to better approximate real-world usage. We also note that memory persistence between conversations varies across LLM providers and models.\footnote{OpenAI (GPT-4o-mini): \url{https://help.openai.com/en/articles/8590148-memory-faq}; Google (Gemini 2.0 Flash): \url{https://blog.google/feed/gemini-referencing-past-chats/}; Anthropic (Claude 3.7 Sonnet): \url{https://support.claude.com/en/articles/11817273-using-claude-s-chat-search-and-memory-to-build-on-previous-context}}

Third, although we examined robustness across different sociodemographic personas, we did not analyze second-order effects---such as how these traits might indirectly influence linguistic style, framing, or topical emphasis in model outputs. Future work could extend this analysis and consider alternative personalization dimensions (e.g., political orientation or domain expertise).

Finally, our comparison relied on Google News as the primary baseline. While suitable for studying media exposure and diversity, including direct Google Search results could provide a more comprehensive benchmark. We leave this for future work, noting that news-related queries in Google Search often redirect to Google News results. We excluded specialized GPTs\footnote{\url{https://chatgpt.com/gpts}}
or custom agents, given their limited adoption and the lack of transparency regarding their underlying system prompts.

\spara{Impact.} %
LLMs increase the opacity of information retrieval. Whereas the first web search engines returned results based on keyword matching and over time they have developed more sophisticated retrieval and ranking mechanisms, the internal processes by which LLMs with web access formulate and execute queries is even more opaque. Recent evidence also suggests a shift in user behavior: a July 2025 Pew Research Center survey found that users are significantly less likely to click on links when an AI-generated summary is presented in search results.\footnote{\url{https://www.pewresearch.org/short-reads/2025/07/22/google-users-are-less-likely-to-click-on-links-when-an-ai-summary-appears-in-the-results/}}
These trends underscore the urgency of transparent auditing and responsible design for LLM-based information systems.

We believe that this study is timely as LLM agents increasingly serve as interfaces for information retrieval, we must avoid repeating the mistakes of past technologies. This study builds on two decades of research in algorithmic auditing, which has shown how search engines can shape public understanding of news and information. Our findings suggest that LLM-mediated retrieval may further narrow the scope of information exposure. Such dynamics could have a compounding effect if LLMs are also used to generate news content—a trend already being explored by several major news organizations, including Politico,\footnote{\url{https://www.wired.com/story/politico-workers-axel-springer-artificial-intelligence/}} The Guardian\footnote{\url{https://www.theguardian.com/gnm-press-office/2025/feb/14/guardian-media-group-announces-strategic-partnership-with-openai}}, and Business Insider.\footnote{\url{https://www.theverge.com/news/779739/business-insider-ai-writing-stories}} Recent policy frameworks, such as the EU Artificial Intelligence Act (AI Act)~\cite{eu2024aiact} and California’s Transparency in Frontier Artificial Intelligence Act (TFAIA)~\cite{ca2024tfaia}, establish explicit transparency and accountability obligations for model providers. These initiatives reflect a growing recognition that large-scale AI systems can exert significant influence on the information ecosystem. In particular, the dissemination and prioritization of news content by LLMs intersect directly with democratic deliberation, electoral processes, and public trust in institutions. The EU AI Act explicitly identifies “systemic risks” arising from foundation models with broad societal impact—including the manipulation of information flows and public opinion—as areas requiring oversight and documentation.

\section{Conclusions}
\label{sec:conclusions}
Our study provides the first systematic audit of how LLM agents retrieve and curate online news.
Across five complementary dimensions—diversity, attention inequality, source composition, ideological orientation, and factual reliability—we find that LLM-mediated news exposure differs markedly from traditional search engines.
Compared to \google{}, LLM agents tend to surface a narrower set of outlets, allocate attention less evenly, and promote distinct mixes of source types and political leanings.
While some systems, notably \gpt{}, elevate outlets with higher factual reliability, others, such as \claude{}, emphasize institutional or non-traditional sources.
These findings highlight that LLMs already enact implicit editorial choices shaped by their underlying architectures and alignment policies.

Our robustness analyses further show that these patterns persist across variations in sociodemographic prompting and when using independent reliability benchmarks such as Wikipedia’s Perennial Sources List.
Taken together, our results underscore that generative models are not neutral conduits of information but active mediators of media visibility.
As LLMs become integrated into everyday search and recommendation tools, understanding—and governing—their editorial behavior will be crucial for ensuring transparency, pluralism, and trust in the digital information ecosystem.

\begin{acks}
MM and GM acknowledge partial support
by $(i)$ the SERICS project (PE00000014) under the NRRP MUR program funded by the EU - NGEU and $(ii)$ European Union - NextGenerationEU - National Recovery and Resilience Plan Project: “SoBig-Data.it - Strengthening the Italian RI for Social Mining and Big Data Analytics” - Prot. IR0000013.
\end{acks}

\bibliographystyle{ACM-Reference-Format}
\bibliography{references}

\clearpage
\appendix
\setcounter{table}{0}        
\renewcommand{\thetable}{A\arabic{table}}
\label{sec:appendix}
\section{Motivation}\label{sec:appendix-motivation}

\spara{Prevalence of News Retrieval Requests in Large LLM Corpora.} %
This study examines how LLM agents function as search engines for news retrieval. Our focus is motivated both by our core research question of understanding LLMs' editorial and retrieval biases, and by the practical observation that a substantial share of real-world interactions involve requests for recent or factual information available on the web.

To contextualize this focus, we estimate the prevalence of news-related prompts in large public chat datasets. We analyze three representative corpora---\textit{WildChat-1M}~\cite{zhao2024wildchat1m}, \textit{WildChat-50M}~\cite{feuer2025wildchat50m}~\cite{zheng2023lmsyschat1m}, and \textit{LMSYS-Chat-1M}--which provide large-scale, openly available records of user-LLM conversations across multiple model families. These datasets reflect organic usage patterns rather than synthetic benchmarks, making them suitable for estimating how often users query LLMs about current events.

For each dataset, we first detected the language of every conversation and retained only English ones. We then applied regular expressions to identify prompts indicative of news-related information seeking. As shown in Table~\ref{tab:llm-news-prompts}, such prompts represent a non-negligible fraction of English-language messages, underscoring the relevance of studying LLMs as news intermediaries. While this estimation is not exhaustive, it provides a lower-bound indication of how frequently users rely on LLMs for news queries. Notably, some conversations predate the introduction of web-search capabilities in these models, meaning that certain responses may consist of generic disclaimers rather than retrieved content. The fact that users where prompting models from news when this functionality was not available yet is another sign of the natural interface that LLM present for news searching.

\begin{table}[h!]
  \centering\small
  \setlength{\tabcolsep}{6pt}
  \begin{tabular}{@{}lrrrr@{}}
\toprule
\textbf{Dataset} &
\multicolumn{1}{l}{\textbf{\makecell{Release\\(mm/yyyy)}}} &
\multicolumn{1}{l}{\textbf{\makecell{English\\E}}} &
\multicolumn{1}{l}{\textbf{\makecell{News\\N}}} &
\multicolumn{1}{l}{\textbf{$\mathbf{\dfrac{E}{N}}$ (\%)}} \\
\midrule
WildChat-1M~\cite{zhao2024wildchat1m}     & 05/2023 & 478,498    & 23,863  & 4.99 \\
WildChat-50M~\cite{feuer2025wildchat50m}  & 01/2025 & 26,620,123 & 608,364 & 2.29 \\
lmsys-chat-1m~\cite{zheng2023lmsyschat1m} & 09/2023 & 777,453    & 10,211  & 1.31 \\
\bottomrule
\end{tabular}

  \caption{Proportion of news-related prompts among English-language prompts.}
  \label{tab:llm-news-prompts}
\end{table}

\section{Dataset}\label{sec:appendix-dataset}
In this section, we describe and motivate the dataset selection procedure. First we describe how we selected the news topics that we have considered and then we describe the sources categorization that we used to augment our dataset.

\spara{Topic Selection.}%
Table~\ref{tab:topics} reports the full list of topics considered.  
We selected topics spanning major domains of sociopolitical relevance, including conflicts, elections, economic policy, public health, climate change, and socio-cultural debates. 
Specifically, our 24 topics were chosen to combine: (i) highly current issues dominating recent news cycles; (ii) polarizing content which, in line with prior work on contested issues and polarization \cite{garimella2018political}, is more likely to expose ideological differences; (iii) operational consistency, achieved by adopting Wikipedia article titles as canonical topic labels to avoid framing bias; and (iv) geographic diversity, ensuring coverage of different international contexts. 
This strategy balances breadth and salience, enabling a systematic assessment of LLM editorial behavior across diverse issue types and regions.

\begin{table}[hb!]
  \centering\small
  \setlength{\tabcolsep}{6pt}    
  \resizebox{\columnwidth}{!}{
\begin{tabular}{p{0.49\linewidth} p{0.49\linewidth}}
  \toprule
  \multicolumn{2}{c}{\textbf{Topics in our audit corpus}}\\
  \midrule
  1. Russian invasion of Ukraine                       & 13. Iran–United States relations      \\
  2. Gaza war                                          & 14. M23 campaign                      \\
  3. Romanian Presidential Election                    & 15. US immigration policy             \\
  4. Poland Presidential Election                      & 16. Abortion policy                   \\
  5. Red Sea crisis                                    & 17. Gender identity                   \\
  6. US tariffs                                        & 18. Gun control policy                \\
  7. Conclave                                          & 19. Marijuana legalization            \\
  8. German AfD ruled extremist                        & 20. Same-sex marriage                 \\
  9. Climate change                                    & 21. LGBTQ rights                      \\
  10. Freedom of the press (US)                        & 22. Inflation                         \\
  11. South-west US measles outbreak                   & 23. Sea-level rise                   \\
  12. India--Pakistan border skirmishes                & 24. Persecution of Uyghurs in China  \\
  \bottomrule
\end{tabular}
}

  \caption{Complete list of the 24 sociopolitical topics used to query each LLM news agent over the audit period.}
  \label{tab:topics}
\end{table}

\spara{Source Categorization.}%
Table~\ref{tab:IAB-categories} presents the main categories defined in the IAB Content Taxonomy 3.0.
Domain classifications were obtained using the third-party service \texttt{Klazify}, which provides a confidence score for each domain. In total, 290 distinct domains were classified: 261 returned valid IAB categories with a confidence score of $\geq 0.50$, while the remaining 30 returned a \texttt{null} classification. The latter were manually categorized following the reference taxonomy. 

\begin{table}[hb!]
  \centering\small
  \setlength{\tabcolsep}{6pt}
  \begin{tabular}{@{}l r l r@{}}
\toprule
  \multicolumn{1}{l}{\textbf{Category}} %
& \multicolumn{1}{l}{\textbf{\# D}} %
& \multicolumn{1}{l}{\textbf{Category}} %
& \multicolumn{1}{l}{\textbf{\# D}} \\
\midrule
Adult                     & 0  & Jobs \& Education        & 6  \\
Arts \& Entertainment     & 25 & Law \& Government        & 19 \\
Autos \& Vehicles         & 1  & News                    & 140 \\
Beauty \& Fitness         & 2  & Online Communities       & 0  \\
Books \& Literature       & 3  & People \& Society        & 18 \\
Business \& Industrial    & 16 & Pets \& Animals          & 0  \\
Computers \& Electronics  & 0  & Real Estate             & 0  \\
Finance                   & 3  & Reference               & 0  \\
Food \& Drink             & 0  & Science                 & 9  \\
Games                     & 0  & Sensitive Subjects       & 7  \\
Health                    & 2  & Shopping                & 2  \\
Hobbies \& Leisure        & 0  & Sports                  & 1  \\
Home \& Garden            & 0  & Travel                  & 3  \\
Internet \& Telecom       & 1  & \texttt{null}           & 32 \\
\bottomrule
\end{tabular}
  \caption{The 27 main categories of the IAB Content Taxonomy 3.0 and the number of domains in our dataset categorized in that class.}
  \label{tab:IAB-categories}
\end{table}

\end{document}